\documentclass[showpacs,twocolumn,preprintnumbers,showkeys,superscriptaddress,amsmath,amssymb,nofootinbib]{revtex4}
\usepackage{graphicx}
\usepackage{color}
\usepackage{bbold}
\usepackage{color}
\usepackage{latexsym}
\usepackage{amsmath}
\usepackage{amssymb}
\usepackage{epstopdf}
\usepackage{epsfig}
\usepackage[utf8]{inputenc}
\usepackage{amsfonts}
\usepackage{bm}
\usepackage{bbold}
\usepackage{amssymb}

\usepackage{eufrak}
\usepackage{euscript}
\usepackage{pstricks}
\usepackage{graphics}
\usepackage{graphicx}
\usepackage{picture}
\begin{document}
\title{ \bf Localization of a $3$-form gauge field in a braneworld}         
\author{Pratik Chattopadhyay} \email{pratikpc@gmail.com}
\affiliation{Department of Physics, Ariel University, Ariel 40700, Israel.}
\author{M. J. Neves} \email{mariojr@ufrrj.br}
\affiliation{Departamento de F\'{i}sica, Universidade Federal Rural do Rio de Janeiro, 
\\
BR 465-07, 23890-971, Serop\'edica, RJ, Brasil.}
%

%
%
%
\begin{abstract}
The localization of a $3$-form gauge field is investigated in a extra dimension attached 
to an AdS metric in $4+1$ dimensions. We propose the study of a model for the scalar field and a $3$-form gauge field 
coupled to the gravitational field in $5D$. The spacetime is set by an AdS metric whose geometry introduces a warped function 
of the fifth spatial coordinate. Considering that the scalar field also depends only on the extra coordinate, we obtain the equation of motions for the set of fields of the model. Using the technique of variable separation in the $3$-rank gauge sector, we explore the localization 
of the dynamical in terms of the warped function. We also extend the initial model for a complex scalar field that couples minimally with the dual 
strength field tensor associated with the $3$-rank gauge field. Thus, the coupling constant has a fundamental role in the localization of 
new $p$-form gauge field on the brane domain.

\end{abstract}


\maketitle

\pagestyle{myheadings}
\markright{Localization of a $3$-form gauge field in a braneworld}

%
%
\section{Introduction}       
The possibility that our four-dimensional universe is embedded in a higher-dimensional spacetime has attracted considerable attention in theoretical physics, particularly in the context of braneworld scenarios \cite{Dvali}. In these models, ordinary matter and gauge fields may be confined to a four-dimensional hypersurface, while gravity propagates in the entire higher-dimensional bulk. A particularly interesting realization is provided by warped geometries, in which the extra spatial dimension is characterized by a nontrivial warp factor. The Randall–Sundrum scenario demonstrated that such geometries can provide an elegant framework for addressing the hierarchy between the gravitational and electroweak scales \cite{Randall}. Furthermore, thick-brane models, in which the brane has a finite thickness and is dynamically generated by scalar fields, provide a natural setting for investigating the localization of matter and gauge fields in higher-dimensional spacetimes \cite{Folomeev}. In theoretical physics, 
superstring theory and quantum gravity are scenarios of extra-dimensions that naturally includes braneworld frameworks \cite{MGreen}.    
An important issue in braneworld physics concerns the localization of gauge fields with different tensorial structures. While the localization mechanism for scalar and gravitational fields has been extensively investigated, antisymmetric tensor fields require special attention because their kinetic terms depend explicitly on the geometry of the extra dimension. As examples, see the rank-$2$ antisymmetric gauge field known as Kalb-Ramond \cite{Kalb}, and the $3$-form gauge field in scenarios of quantum chromodynammics (QCD) \cite{Aurilia,Hata,Dvali1,DvaliPRD}. In particular, higher-rank antisymmetric gauge fields, described by differential forms, play an interesting role in higher-dimensional field theories and may lead to nontrivial effective interactions after dimensional reduction. The dynamics and duality properties of these fields have been investigated in several contexts \cite{Marc,Nitta,Anber}. In five dimensions, a rank-three gauge potential is associated with a rank-four field-strength tensor \cite{Lusher}, whose propagation and effective four-dimensional dynamics are sensitive to the warped geometry.
The localization of antisymmetric tensor fields on thick branes has been studied in different gravitational backgrounds, revealing that the behavior of their effective modes depends strongly on the warp factor and on the structure of the corresponding kinetic terms \cite{CruzEPL2009,CruzPRD2013,CruzEPJC2013}. In this context, the separation of variables provides a useful method for investigating the dependence of the higher-dimensional gauge field on the extra coordinate and for determining the conditions under which its effective four-dimensional degrees of freedom can be localized. Moreover, dimensional reduction can generate effective mass terms and modify the kinetic structure of the lower-dimensional theory. These features motivate a systematic analysis of higher-rank gauge fields in warped five-dimensional geometries.
In this work, we investigate the model of a scalar field and an antisymmetric rank-3 gauge potential coupled to gravity in five dimensions. We consider a warped anti-de Sitter geometry in which the warp factor depends exclusively on the extra spatial coordinate. The scalar field is also assumed to depend only on this coordinate, allowing us to obtain the corresponding gravitational and scalar field equations and to investigate their solutions in a thick-brane background. The field-strength tensor associated with the rank-three gauge potential is then analyzed by employing a separation-of-variables approach. In particular, we derive the differential equation governing its extra-dimensional profile and investigate the resulting effective four-dimensional action. This analysis allows us to distinguish the contributions associated with the kinetic and mass terms of the reduced gauge field and to examine the localization properties of the corresponding modes.
Posteriorly, we extend the model by introducing a complex scalar field minimally coupled to both an electromagnetic gauge potential and the dual field-strength tensor associated with the rank-three gauge potential. The generalized covariant derivative introduces a new interaction controlled by an additional coupling constant. This extension modifies the equation of motion of the antisymmetric gauge field and, consequently, changes the differential equation governing its extra-dimensional profile. The resulting dimensional reduction generates additional contributions to the effective four-dimensional action. In particular, the coupling to the dual field-strength tensor gives rise to a new kinetic contribution associated with an effective Kalb-Ramond sector. Its dependence on the extra-dimensional coordinate provides a mechanism through which the corresponding effective field can become localized around the brane.
The main purpose of this work is the investigation of the minimal coupling between a complex scalar field and the dual field-strength tensor of a rank-$3$ gauge field affects the localization mechanism in a warped five-dimensional spacetime. By comparing the uncoupled and coupled configurations, we identify the role played by the additional coupling constant in modifying the extra-dimensional profiles and the effective four-dimensional dynamics. Our results indicate that interactions involving higher-rank gauge fields and scalar matter can provide new mechanisms for generating localized gauge degrees of freedom in thick-brane scenarios.
The paper is organized as follows. In Section \ref{sec2}, we introduce the five-dimensional gravitational model containing a scalar field and a rank-three antisymmetric gauge potential. We derive the corresponding field equations, consider the warped metric and investigate the localization properties of the gauge field through dimensional reduction. In Section \ref{sec3}, we extend the model by introducing a complex scalar field minimally coupled to the electromagnetic potential and to the dual field-strength tensor of the rank-three gauge field. We derive the modified field equations and analyze the resulting effective four-dimensional action, emphasizing the emergence and localization properties of the Kalb--Ramond sector. Finally, in Section \ref{sec4}, we present our conclusions and discuss the implications of the results.


We use the natural units of $\hbar=c=1$ throughout this paper, with the Minkowski metric of signature $\eta^{\mu\nu}=(-1,+1,+1,+1)$.
In five dimensions, we adopt the hat notation $\hat{\mu}=\left\{ \, 0 \, , \, 1 \, , \, 2 \, , \, 3 \, , \, 4 \, \right\}$ for the 
space-time coordinates in $5D$. 

\section{The gravitational model of a 3-rank field in five dimension}
\label{sec2}
The action of an antisymmetric 3-rank field $C^{\widehat{\mu}\widehat{\nu}\widehat{\kappa}}$ coupled to a gravitational background
in five dimensions is set by
\begin{eqnarray}\label{action5D}
S_{5D} \!&=&\! \int d^{5}X \, \sqrt{-g} \, \left[ \, 2M^3 R-\frac{1}{2} \, g_{\hat{\mu}\hat{\nu}}(X)\partial^{\hat{\mu}}\phi\,\partial^{\hat{\nu}}\phi
-V(\phi)
\right.
\nonumber \\
&&
\left.
\hspace{-0.5cm}
-\frac{1}{48} \, H_{\hat{\mu}\hat{\nu}\hat{\kappa}\hat{\lambda}}\,H^{\hat{\mu}\hat{\nu}\hat{\kappa}\hat{\lambda}} \, \right] , \;\;
\end{eqnarray}
where the hat index means $\hat{\mu}=(\mu,4)$, the coordinates notation is $X^{\hat{\mu}}=(x^{\mu},y)$, in which
$y$ sets the fifth dimension, $M=1.22 \times 10^{19} \, \mbox{GeV}$ is the Planck mass, $R=g_{\hat{\mu}\hat{\nu}}\,R^{\hat{\mu}\hat{\nu}}$ 
is the Ricci scalar, and $H_{\hat{\mu}\hat{\nu}\hat{\kappa}\hat{\lambda}}$ is the field strength tensor of the rank-3 field \cite{Lusher}. 
The form of $H_{\hat{\mu}\hat{\nu}\hat{\kappa}\hat{\lambda}}$ is read as 
\begin{equation}\label{tensorH}
H_{\hat{\mu}\hat{\nu}\hat{\kappa}\hat{\lambda}}=\partial_{\hat{\mu}}C_{\hat{\nu}\hat{\kappa}\hat{\lambda}}
+\partial_{\hat{\nu}}C_{\hat{\lambda}\hat{\kappa}\hat{\mu}}+\partial_{\hat{\kappa}}C_{\hat{\lambda}\hat{\mu}\hat{\nu}}
-\partial_{\hat{\lambda}}C_{\hat{\mu}\hat{\nu}\hat{\kappa}} \; ,
\end{equation}
with $C_{\hat{\nu}\hat{\kappa}\hat{\lambda}}$ being the rank-3 gauge field totally antisymmetric, that obeys the gauge transformation
\begin{eqnarray}\label{transfC}
C^{\,\prime}_{\hat{\mu}\hat{\nu}\hat{\rho}}=C_{\hat{\mu}\hat{\nu}\hat{\rho}}
+\partial_{\hat{\mu}}\xi_{\hat{\nu}\hat{\rho}}+\partial_{\hat{\nu}}\xi_{\hat{\rho}\hat{\mu}}
+\partial_{\hat{\rho}}\xi_{\hat{\mu}\hat{\nu}} \; ,
\end{eqnarray}
where $H_{\hat{\mu}\hat{\nu}\hat{\kappa}\hat{\lambda}}$ is invariant for an antisymmetric and arbitrary tensor $\xi_{\hat{\mu}\hat{\nu}}=-\xi_{\hat{\nu}\hat{\mu}}$.
The derivative notation in five dimensions means the derivation in relation to $X^{\hat{\mu}}$, {\it i. e.},
$\partial_{\hat{\mu}}\equiv \frac{\partial}{\partial X^{\hat{\mu}}}$. The scalar field $\phi$ is minimally 
coupled to gravitational field in which a $V(\phi)$-potential is so included and it plays an important role in the model.
Using the action principle for the rank-3 gauge field, the correspondent field equation is
\begin{eqnarray}\label{EqmotionH}
\nabla_{\hat{\mu}} H^{\hat{\mu}\hat{\nu}\hat{\rho}\hat{\sigma}}=0 \; ,
\end{eqnarray}
where $\nabla_{\hat{\mu}}$ is the covariant derivative operator in the curved $5D$ spacetime. 
The action principle applied to the scalar field yields the field equation
\begin{eqnarray}\label{Eqscalar}
\partial^{\hat{\nu}}\left[ \, \sqrt{-g} \, g_{\hat{\mu}\hat{\nu}}(X) \, \partial^{\hat{\mu}}\phi \, \right]-\sqrt{-g}\,\frac{\partial V}{\partial \phi}=0 \; .
\end{eqnarray}
The gravitational equation field is
\begin{eqnarray}\label{EqEinstein5D}
G_{\hat{\mu}\hat{\nu}}
= \frac{1}{4M^{3}} \, T_{\hat{\mu}\hat{\nu}} \; ,
\end{eqnarray}
where $G_{\hat{\mu}\hat{\nu}}=R_{\hat{\mu}\hat{\nu}}-\frac{1}{2}\, g_{\hat{\mu}\hat{\nu}} R$ 
is the Einstein tensor, $T_{\hat{\mu}\hat{\nu}}=T_{\hat{\mu}\hat{\nu}}^{(\phi)} + \, T_{\hat{\mu}\hat{\nu}}^{(C)}$ 
is the total energy-momentum tensor of the model, in which $T_{\hat{\mu}\hat{\alpha}}^{(\phi)}$ is associated with the scalar field, 
and $T_{\hat{\mu}\hat{\nu}}^{(C)}$ is the 3-rank energy-momentum tensor. Namely, these tensors are read below :
\begin{subequations}
\begin{eqnarray}
T_{\hat{\mu}\hat{\nu}}^{(\phi)} &=& \partial_{\hat{\mu}}\phi\,\partial_{\hat{\nu}}\phi
-g_{\hat{\mu}\hat{\nu}}\left[ \, \frac{1}{2} \, g_{\hat{\rho}\hat{\lambda}}(X) \, \partial^{\hat{\rho}}\phi \, \partial^{\hat{\lambda}}\phi+V(\phi) \, \right] ,
\hspace{0.8cm}
\\
T_{\hat{\mu}\hat{\nu}}^{(C)} &=& 
\frac{1}{6} \, H_{\hat{\mu}\hat{\rho}\hat{\kappa}\hat{\lambda}}\,
H_{\hat{\nu}}^{\;\;\,\hat{\rho}\hat{\kappa}\hat{\lambda}}
-g_{\hat{\mu}\hat{\nu}} \, \frac{1}{48} \, 
\,H_{\hat{\rho}\hat{\kappa}\hat{\lambda}\hat{\sigma}}
H^{\hat{\rho}\hat{\kappa}\hat{\lambda}\hat{\sigma}} \; .
\end{eqnarray}
\end{subequations}
Since the Einstein tensor satisfies the identity $\nabla_{\hat{\mu}}G^{\hat{\mu}\hat{\nu}}=0$ 
in five dimensions, the field equation (\ref{EqEinstein5D}) leads to the 
conservation of the total energy-momentum tensor $\nabla_{\hat{\mu}}T^{\hat{\mu}\hat{\nu}}=0$. 
%

%
The line element of this spacetime is set by
\begin{equation}\label{lineelement}
ds^{2}=g_{\hat{\mu}\hat{\nu}}(X) \, dX^{\hat{\mu}}\,dX^{\hat{\nu}}=e^{2A(y)} \, \eta_{\mu\nu} \, dx^{\mu}\,dx^{\nu}+dy^{2} \; ,
\end{equation}
where $\eta_{\mu\nu}=\mbox{diag}(-1,+1,+1,+1)$ is the Minkowski spacetime metric, and $A(y)$ is a function
of the fifth $y$-coordinate. In this approach, the $\phi$-scalar field depends only on the coordinate $y$, thereby,
the scalar equation (\ref{Eqscalar}) is reduced to :
\begin{eqnarray}\label{Eqphi}
\frac{d^2\phi}{d y^2}+4\,\frac{d A}{dy}\frac{d\phi}{dy}-\frac{\partial V}{\partial \phi} = 0 \; .
\end{eqnarray}
In the absence of the $3$-form, the Einstein equations (\ref{EqEinstein5D}) lead to two equations
\begin{subequations}
\begin{eqnarray}
\frac{d^2A}{dy^2} &=& \frac{-1}{12M^3} \left(\frac{d\phi}{dy}\right)^{2} \; ,
\label{Eqd2A}
\\
\left( \frac{dA}{dy} \right)^2 &=& \frac{1}{24M^3}\left[ \, \frac{1}{2} \, \left(\frac{d\phi}{dy}\right)^{2}-V(\phi)  \, \right] \; .
\label{EqdA}
\end{eqnarray}
\end{subequations}
%
%

%
%
%
%
The field equation for the rank-3 (\ref{EqmotionH}) for the metric (\ref{lineelement}) is written 
as
\begin{equation}\label{Eqmotion4+1}
e^{-2A(y)} \, \partial_{\mu}H^{\mu\nu\kappa\lambda}-\partial_{y}H^{\nu\kappa\lambda4}+2 \left( \frac{dA}{dy} \right) H^{\nu\kappa\lambda4}=0 \; .
\end{equation}
Making the variable separation
\begin{subequations}
\begin{eqnarray}
C_{\hat{\mu}\hat{\nu}\hat{\kappa}}(X) &=& c_{\hat{\mu}\hat{\nu}\hat{\kappa}}(x) \, U(y) \; ,
\label{Cmunukappasplit}
\\
H_{\hat{\mu}\hat{\nu}\hat{\kappa}\hat{\lambda}}(X) &=& h_{\hat{\mu}\hat{\nu}\hat{\kappa}\hat{\lambda}}(x) \, U(y) \, ,
\label{Hmunukappalambdasplit}
\end{eqnarray}
\end{subequations}
and choosing the gauge condition $\partial_{\mu}C^{\mu\nu\kappa}=C^{\nu\kappa4}=0$, 
the $H_{\mu\nu\kappa 4}$-component is reduced to
\begin{eqnarray}\label{Hmunukappay}
H_{\mu\nu\kappa 4}=-c_{\mu\nu\kappa}(x) \, \frac{dU}{dy} \; .
\end{eqnarray}
Hence one must to solve the equation motion (\ref{Eqmotion4+1}) for the function $U$, in the case of a
massless field $c_{\mu\nu\kappa}$ with the differential equation
\begin{eqnarray}\label{EqdiffU}
\frac{d}{dy}\left[ \, e^{-2A(y)} \, \frac{dU}{dy} \, \right]=0 \; ,
\end{eqnarray}
that has the following solution
\begin{eqnarray}\label{solutionU}
U(y)=C_{1} \int_{y} \, dy^{\prime} \, e^{2A(y^{\prime})}+C_{2} \; ,
\end{eqnarray}
where $C_{1}$ and $C_{2}$ are real constants, and we can obtain the solution for the $U(y)$-function in terms of $A(y)$.
Considering the scalar potential for a quartic self-interaction
\begin{eqnarray}\label{potential}
V(\phi)=\frac{\lambda}{4} \left(\phi^{2}-v^{2}\right)^{2} \; ,
\end{eqnarray}
where $v$ is a vacuum expectation value (VEV), and $\lambda$ is a real coupling constant, the solutions for $\phi(y)$ and $A(y)$ are given by
\begin{subequations}
\begin{eqnarray}\label{phiy}
\phi(y) &=& v \; \mbox{tanh}(ay) \; ,
\label{phiy}
\\
A(y) &=& -\,\beta \, \ln \mbox{cosh}(ay)^2-\frac{\beta}{4} \, \mbox{tanh}^2(ay) \; ,
\label{Ay}
\end{eqnarray}
\end{subequations}
where we have defined the positive constants $a^2=\lambda v^{2}/2$, and $\beta=v^{2}/(18M^{3})$. 

\begin{figure}[h]
\centering
\includegraphics[width=0.9\linewidth]{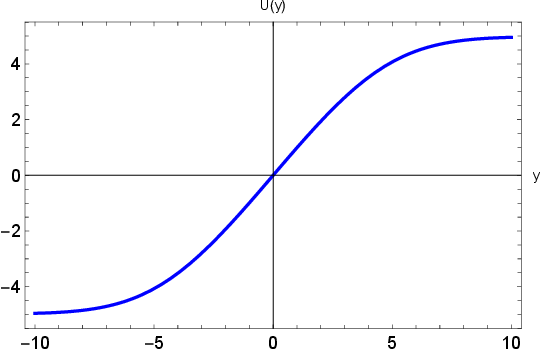}
\caption{The $U(y)$ as function of the $y$-coordinate for $C_1=1$, $C_2=0$, $a=0.5$ (in mass squared units) and $\beta=0.1$ (in length dimension).}\label{fig1}
\end{figure}

The numerical plot for the $U(y)$-function is illustrated in the figure (\ref{fig1}). In this figure, 
we use the values of $C_1=1$, $C_2=0$, with $a=0.5$ (in mass squared units) and $\beta=0.1$ (in length dimension). 
%

%
%
%
%

\begin{figure*}[th]
\centering
\includegraphics[width=0.43\textwidth]{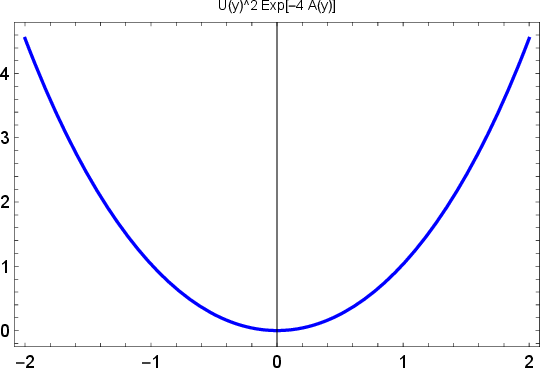} \quad\quad
\includegraphics[width=0.44\textwidth]{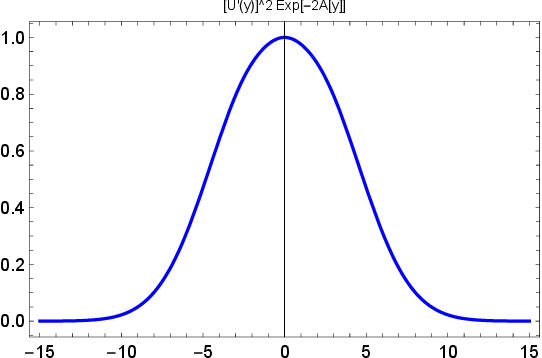}
\caption{ Left panel : The function $U(y)^2\,e^{-4A(y)}$ versus $y$-coordinate from kinetic term of the $3$-rank. Right panel : 
    The function $\left(dU/dy\right)^{2} e^{-2A(y)}$ versus $y$ for the massive term of $3$-rank. } 
\label{fig2}
\end{figure*}

Using the variable separation (\ref{Cmunukappasplit})-(\ref{Hmunukappalambdasplit}), we can split the action (\ref{action5D}) 
into $4+1$ dimensions as
\begin{eqnarray}\label{SepAction}
S_{5D} &=& -\frac{1}{48}\int_{-\infty}^{\infty} dy \, U^{2}(y) \, e^{-4A(y)} \int d^{4}x \, h_{\mu\nu\kappa\lambda}^{2}
\nonumber \\
&&
\hspace{-0.5cm}
-\frac{1}{12}\int_{-\infty}^{\infty} \left(\frac{dU}{dy} \right)^{2} e^{-2A(y)}\int d^{4}x \; c_{\mu\nu\kappa}^{2} \; ,
\end{eqnarray}
in which the extra dimension through the functions $A(y)$ and $U(y)$ have a fundamental role in the location of the tensor field inside the membrane. 
The first term contains to the kinetic term of a $3$-form gauge in four dimensions, while that the second term is interpreted as a massive term for the 
$3$-form attached to extra-dimension. Thereby, we plot the functions $U^{2}(y) \, e^{-4A(y)}$ and $\left(\frac{dU}{dy} \right)^{2} e^{-2A(y)}$ in the 
figure (\ref{fig2}). 
%

%

%

The behavior
divergent of those integral lead us to interpretation that the rank-$3$ antisymmetric field is not localized in the membrana. Nevertheless the second term generates a mass for $c_{\mu\nu\kappa}$, because we have fixed a gauge, in that the $y$-integral is well defined.
The graphics of those integrals are illustrated above
%
%


%
%
%
%
\section{The complex scalar field minimally coupled to a 3-rank gauge in five dimensions}
\label{sec3}
An extension of the model discussed in the previous is the inclusion of the minimal coupling of the rank-3 gauge field with 
a complex scalar field. This model is set by the action 
\begin{eqnarray}\label{action5DH}
S_{5D} &=& \int d^{5}X \sqrt{-g} \left[ \phantom{\frac{1}{2}} \!\!\!\! 2M^3 R
\right. 
\nonumber \\
&&
\left. 
\hspace{-0.8cm}
-\frac{1}{2} \, g_{\hat{\mu}\hat{\nu}}(X) 
(D^{\hat{\mu}}\phi)^{\ast} D^{\hat{\nu}}\phi-V(\phi^{\ast},\phi)
\right. 
\nonumber \\
&&
\left. 
\hspace{-0.8cm}
-\frac{1}{4} \, F_{\hat{\mu}\hat{\nu}}F^{\hat{\mu}\hat{\nu}}
-\frac{1}{48} \, 
H_{\hat{\mu}\hat{\nu}\hat{\kappa}\hat{\lambda}}
H^{\hat{\mu}\hat{\nu}\hat{\kappa}\hat{\lambda}} \, \right] \, , \hspace{0.5cm}
\end{eqnarray}
where the covariant derivative operator is 
\begin{eqnarray}\label{DcovH}
D_{\hat{\mu}}=\partial_{\hat{\mu}} \, + i \, e \, A_{\hat{\mu}} + \, i \, g_{0} \, H_{\hat{\mu}} \; ,
\end{eqnarray}
that couples the complex scalar field to the dual field strength tensor $H_{\hat{\mu}}=\epsilon_{\hat{\mu}\hat{\nu}\hat{\kappa}\hat{\lambda}\hat{\rho}}
H^{\hat{\nu}\hat{\kappa}\hat{\lambda}\hat{\rho}}/24$, $g_{0}$ being 
a constant coupling with length dimension, and with the EM $A^{\hat{\mu}}$-potential associated with the field strength tensor $F^{\hat{\mu}\hat{\nu}}=\partial^{\hat{\mu}}A^{\hat{\nu}}-\partial^{\hat{\nu}}A^{\hat{\mu}}$ in 5D. 
The action (\ref{action5DH}) exhibits a $U(1)$-local gauge 
invariance with $\phi \mapsto \phi^{\prime} (X) = e^{i\, \alpha(X)} \, \phi(X)$, in which 
the rank-3 gauge field transforms 
under (\ref{transfC}), with $\alpha(X)$ an arbitrary function of $X^{\hat{\mu}}$, and $A^{\prime\hat{\mu}}=A^{\hat{\mu}}-\partial^{\hat{\mu}}\alpha$. 
The scalar potential is modified to     
\begin{eqnarray}\label{potentialphicomplex}
V(\phi^{\ast},\phi)=\frac{\lambda}{4} \left( \, \phi^{\ast}\,\phi-v^{2} \,\right)^{2} \; .
\end{eqnarray}
%

\begin{figure}[h]
\centering
\includegraphics[width=0.99\linewidth]{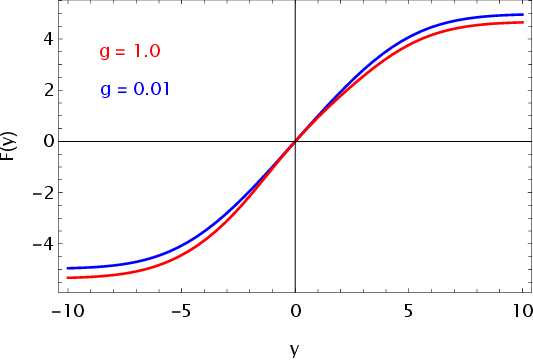}
\caption{The $F(y)$ as function of the $y$-coordinate for $C^{\prime}_1=1$, $C^{\prime}_2=0$, $a=0.5$ (in mass squared units), $\beta=0.1$ (in length dimension), $g_{0}=0.01$ (blue line) and $g_{0}=1.0$ (red line).}\label{Fy}
\end{figure}

%
%
The electromagnetic field equation associated with the action (\ref{action5DH}) is set by 
\begin{eqnarray}
\nabla_{\hat{\mu}}F^{\hat{\mu}\hat{\nu}}=-\frac{ie}{2} \left[ \, \phi^{\ast} D^{\hat{\nu}}\phi - (D^{\hat{\nu}}\phi)^{\ast} \phi \, \right] \; ,
\end{eqnarray}
and the equations without sources are described by the Bianchi identity $\nabla_{\hat{\mu}}F_{\hat{\nu}\hat{\rho}}+\nabla_{\hat{\nu}}F_{\hat{\rho}\hat{\mu}}+\nabla_{\hat{\rho}}F_{\hat{\mu}\hat{\nu}}=0$.
The field equation for the 3-rank field is modified by
\begin{equation}\label{EqmotionHdual}
\nabla_{\hat{\mu}}\left[ H^{\hat{\mu}\hat{\nu}\hat{\rho}\hat{\lambda}}
+\epsilon^{\hat{\kappa}\hat{\mu}\hat{\nu}\hat{\rho}\hat{\lambda}} \, \frac{ig_{0}}{2} \left( \phi^{\ast}D_{\hat{\kappa}}\phi- \left(D_{\hat{\kappa}}\phi\right)^{\ast} \phi \right) \right]=0 \; ,
\end{equation}
and the dual strength field tensor satisfies $\nabla_{\hat{\mu}}H^{\hat{\mu}}=0$.
Substituting the metric (\ref{lineelement}) in (\ref{EqmotionHdual}), we obtain
\begin{eqnarray}\label{Eqmotion4+1H}
\left[\,e^{-4A(y)}+\frac{g_{0}^2}{2} \, e^{4A(y)}\,\right]\phi(y)\,\partial_{\mu}H^{\mu\nu\kappa\lambda}+
\nonumber \\
+\partial_{y}\left[\, e^{-2A(y)} \, H^{\nu\kappa\lambda4}
+\frac{g_{0}^2}{2} \, e^{6A(y)} \, \phi(y) \, H^{\nu\kappa\lambda4} \, \right]=0 \; . \hspace{0.5cm}
\end{eqnarray}
Making a similar variable separation as in (\ref{Cmunukappasplit})-(\ref{Hmunukappalambdasplit}), {\it i.e.}, 
\begin{subequations}
\begin{eqnarray}
C_{\hat{\mu}\hat{\nu}\hat{\kappa}}(X) &=& c_{\hat{\mu}\hat{\nu}\hat{\kappa}}(x) \, F(y) \; ,
\label{CmunukappasplitF}
\\
H_{\hat{\mu}\hat{\nu}\hat{\kappa}\hat{\lambda}}(X) &=& h_{\hat{\mu}\hat{\nu}\hat{\kappa}\hat{\lambda}}(x) \, F(y) \, ,
\label{HmunukappalambdasplitF}
\end{eqnarray}
\end{subequations}
with the gauge condition $\partial_{\mu}C^{\mu\nu\kappa}=C^{\nu\kappa4}=0$, we obtain the expression for the 
$F(y)$-function in terms of $A(y)$ and $\phi(y)$
\begin{eqnarray}\label{solutionUH}
F(y)=C^{\prime}_{1} \int_{y} dy^{\prime} \, \frac{e^{2A(y^{\prime})}}{1+\frac{g_{0}^2}{2} \, e^{8A(y^{\prime})} \, \phi(y^{\prime})}+C_{2}^{\prime} \; ,
\end{eqnarray}
where $C_{1}^{\prime}$ and $C_{2}^{\prime}$ are real constants. In the limit $g_{0} \rightarrow 0$, this result reduces to 
$U(y)$-function from the previous section. The $F(y)$-function is plotted in the figure (\ref{Fy}). 
For simplicity, we choose $C_{1}^{\prime}=1$ and $C_{2}^{\prime}=0$. We use the values of $g_{0}=0.01$ (blue line) and $g_{0}=1.0$ (red line) both in length units.    

\begin{figure*}[ht]
\centering
\includegraphics[width=0.30\textwidth]{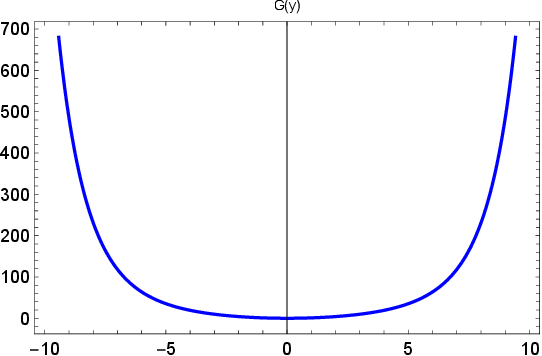} 
\hfill
\includegraphics[width=0.30\textwidth]{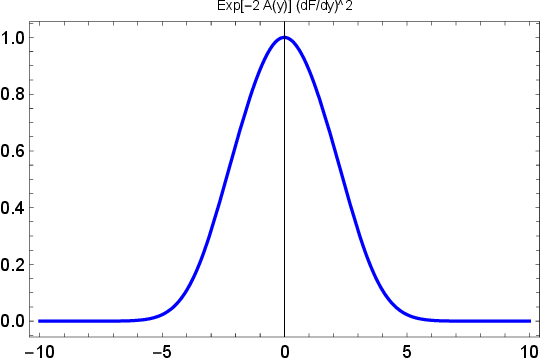} 
\hfill
\includegraphics[width=0.30\textwidth]{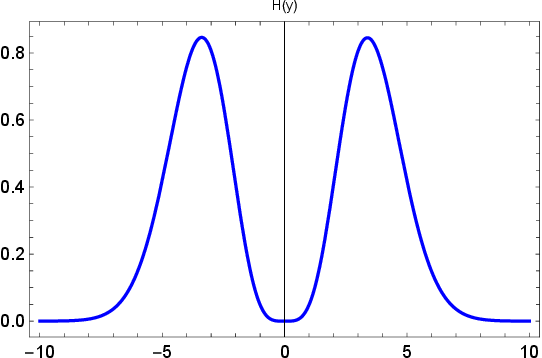}
\caption{Left panel : The function $G(y)$ from eq. (\ref{Gy}) versus the $y$-spatial coordinate. Middle panel : The function 
$\left(dF/dy\right)^{2} e^{-2A(y)}$ versus $y$. Right panel : 
The function $H(y)$ from (\ref{Hy}) versus $y$ for the massive term of $3$-rank in (\ref{ActionSep}).} 
\label{fig4}
\end{figure*}

The separation of the 5D spacetime split in $4+1$ directly into the action (\ref{action5DH}) provides the following terms :
\begin{eqnarray}\label{ActionSep}
S_{5D} &=&-\frac{1}{48}\int_{-\infty}^{\infty} dy \, G(y) \, \int d^{4}x \, h_{\mu\nu\kappa\lambda}^{2} 
\nonumber \\
&&
-\frac{1}{12}\int_{-\infty}^{\infty}dy \, e^{-2A(y)} \, \left(\frac{dF}{dy}\right)^{2}
\int d^{4}x \; c_{\mu\nu\kappa}^{2}
\nonumber \\
&&
+\frac{g_{0}^2}{12} \int_{-\infty}^{\infty}dy \, H(y) \, \int d^{4}x \, G_{\mu\nu\kappa}^{2} \, ,
\;\;\;
\end{eqnarray}
where the functions $G(y)$ and $H(y)$ are, respectively, 
defined by
\begin{subequations}
\begin{eqnarray}
G(y) &=& F^{2}(y) \, \left[ \, e^{-4A(y)}+\frac{g_{0}^2}{24} \, \phi^{2}(y) \, e^{4A(y)} \, \right] \; , \;\;
\label{Gy}
\\
H(y) &=& e^{6A(y)} \, \phi^{2}(y) \, F^{2}(y) \; .
\label{Hy}
\end{eqnarray}
\end{subequations}
The third term in (\ref{ActionSep}) is the new result that depends on the $g_{0}$-coupling. It shows the emergence of a kinetic term for a massless Ralb-Ramond 
field due to the dimensional reduction in the minimal coupling term of the rank-$3$ with the scalar complex field. In the limit $g_{0} \rightarrow 0$, this term 
disappears automatically. The three functions that appear in the integrating from (\ref{ActionSep}) are illustrated in the figure (\ref{fig4}). All the plots 
were generated with the values of $C^{\prime}_{1}=1$, $C^{\prime}_{2}=0$, $\beta=0.1$, $a=0.5$ (length units) and $v=0.5$ (energy units), when $g_{0}=0.1$.  
The $G(y)$-curve (left panel in (\ref{fig4})) shows the divergent behaviour for large $y$-values, that does not guarantee the existence 
of the zero mode for the dynamical of the $3$-rank gauge tensor. The middle plot in (\ref{fig4}) shows a resonance peak for a massive term associated with 
the $3$-rank gauge field. The right panel in (\ref{fig4}) provides two resonance peaks at $y \sim \pm \, 3.34$, that indicates the localization of a 
massless Kalb-Ramond field on the braneworld. 
%


%

\section{conclusions}
\label{sec4}
In this paper, we study the model of a $3$-rank gauge and a scalar field both coupled to the gravitational sector in $5D$ in which the spacetime 
is embedded by a AdS metric governed by a warped factor that is function only of the fifth dimension. We obtain the correspondent field equations
for the scalar field and the warped function in terms of the fifth coordinate, whose solutions we explore 
to determine the localization of the $3$-rank gauge field on the brane-world. In the sector of the $3$-rank, we use a separation variable method to 
introduce a new function of the fifth dimension that is so determined in terms of the warped function defined previously in the AdS metric. 
From original action split in $4+1$ dimensions, we calculate numerically the contributions for the dynamical of $3$-rank, and for a 
massive term that emerge from dimensional reduction. We find a resonance for this mass term due to fifth dimension, 
where the integral in the action is well defined. Posteriorly, we extend the model for a complex scalar field that couple minimally with the 
electromagnetic, and with the $3$-rank gauge field through the dual strength field tensor in $5D$. Using the same metric for the AdS space-time, 
we re-obtain the modified field equation for the $3$-rank gauge field, now under the influence of the coupling constant with the complex scalar field. 
Consequently, the solution for the localization of the $3$-rank is altered, and we investigate the dimensional separation for the action in $5D$. 
The main result is the emergence of a Kalb-Ramond gauge field whose dynamics depends directly on the minimal coupling with the complex scalar field, 
in which this new term is localized on the brane due to a resonance peak well defined in the extra dimension.

\pagebreak

\section*{Acknowledgements}

This work was supported by the Conselho Nacional de Desenvolvimento Cient\'ifico e Tecnol\'ogico (CNPq) under grant 305743/2026-0 (GM) (M. J. Neves).

\end{document}